\documentclass{article}
\usepackage[T1]{fontenc}
\usepackage[utf8]{inputenc}
\usepackage{spconf,amsmath,amsfonts,graphicx,bm}
\usepackage{cite}
\usepackage{microtype}
\usepackage{booktabs}
\usepackage{tabularx}
\usepackage{multirow}
\usepackage[hidelinks]{hyperref}

\allowdisplaybreaks

\title{A DEEP GENERATIVE MODEL OF SPEECH COMPLEX SPECTROGRAMS}

\name{Aditya Arie Nugraha$^{\star}$ \qquad Kouhei Sekiguchi$^{\dagger \star}$ \qquad Kazuyoshi Yoshii$^{\dagger \star}$}

\address{$^{\star}$ RIKEN Center for Advanced Intelligence Project (AIP), Japan \\
	     $^{\dagger}$ Graduate School of Informatics, Kyoto University, Japan}

\begin{document}
\def\baselinestretch{.98}\normalsize
\setlength{\abovedisplayskip}{4pt plus 1.0pt minus 1.0pt}
\setlength{\belowdisplayskip}{4pt plus 1.0pt minus 1.0pt}
\setlength{\abovedisplayshortskip}{4pt plus 1.0pt minus 1.0pt}
\setlength{\belowdisplayshortskip}{4pt plus 1.0pt minus 1.0pt}
\maketitle
\begin{abstract}
This paper proposes an approach to the joint modeling of the short-time Fourier transform magnitude and phase spectrograms with a deep generative model. We assume that the magnitude follows a Gaussian distribution and the phase follows a von Mises distribution. To improve the consistency of the phase values in the time-frequency domain, we also apply the von Mises distribution to the phase derivatives, i.e., the group delay and the instantaneous frequency. Based on these assumptions, we explore and compare several combinations of loss functions for training our models. 
Built upon the variational autoencoder framework, our model consists of three convolutional neural networks acting as an encoder, a magnitude decoder, and a phase decoder.
In addition to the latent variables, we propose to also condition the phase estimation on the estimated magnitude. Evaluated for a time-domain speech reconstruction task, our models could generate speech with a high perceptual quality and a high intelligibility.
\end{abstract}
\vspace{-.75\baselineskip}
\section{Introduction}
\label{sec:intro}
\vspace{-.5\baselineskip}

Speech signal processing methods typically work in the time-frequency (TF) domain, and the most widely used TF representation is the short-time Fourier transform (STFT) \cite{oppenheim09,cohen10,vincent18,allen77assp}. The complex-valued STFT coefficients are typically decomposed into the real-valued magnitude and phase spectrograms.
Most signal processing methods focus on magnitude modification or estimation. However, an increasing number of works have shown that phase, including its derivatives, is useful to improve the performance of various applications \cite{gerkmann15spm,mowlaee16specom}. In this paper, we are interested in the problem of joint magnitude and phase estimation in the context of speech enhancement.

There exists works on phase recovery given the magnitude, including the consistency-based approach \cite{griffin84tassp,leroux08sapa}, the sinusoidal signal model based approaches \cite{magron18taslp,magron18iwaenc}, and the deep neural network (DNN) based approaches \cite{takamichi18iwaenc,takahashi18interspeech}.
Takamichi et al. \cite{takamichi18iwaenc} optimize the estimations of the phase and the group delay assuming a von Mises distribution for each of them. The method outperforms the Griffin-Lim algorithm \cite{griffin84tassp} given the true magnitude. Takahashi et al. \cite{takahashi18interspeech} discretize the phase and view the phase estimation as a classification problem. The method performs well for source separation tasks given the true magnitude and the imperfect magnitude estimate. Both approaches train the DNNs in a supervised manner.

In contrast, a generative model assumes that some observation is generated by some latent variables, and the model learns those variables in an unsupervised manner. Several DNN-based generative models have been proposed recently, including the variational autoencoder (VAE) \cite{kingma14iclr}, the generative adversarial network \cite{goodfellow14nips}, the flow-based model \cite{rezende15icml}, and the autoregressive model, such as the WaveNet \cite{vandenoord16arxiv}.
Among these models, the VAE is arguably the most popular. It has been used for various purposes, including speech separation \cite{kameoka18arxiv,seki18arxiv} and speech enhancement \cite{bando18icassp,leglaive18mlsp,sekiguchi18apsipa}. In these probabilistic approaches, the VAEs act as priors that allow an efficient estimation of the source power spectrograms. We could then employ a phase recovery method to estimate the source phase spectrograms.
However, this cascading approach, i.e., a magnitude estimation followed by a phase recovery, is considered to be suboptimal. Therefore, we aim for a prior of the complex spectrogram. The WaveNet is possibly used to provide a time-domain prior.
Nonetheless, we opt to work in the TF domain so that we can build upon the various speech enhancement methods that work in this domain \cite{vincent18}.

\begin{figure}[!t]
	\centering
	\centerline{\includegraphics[width=\columnwidth]{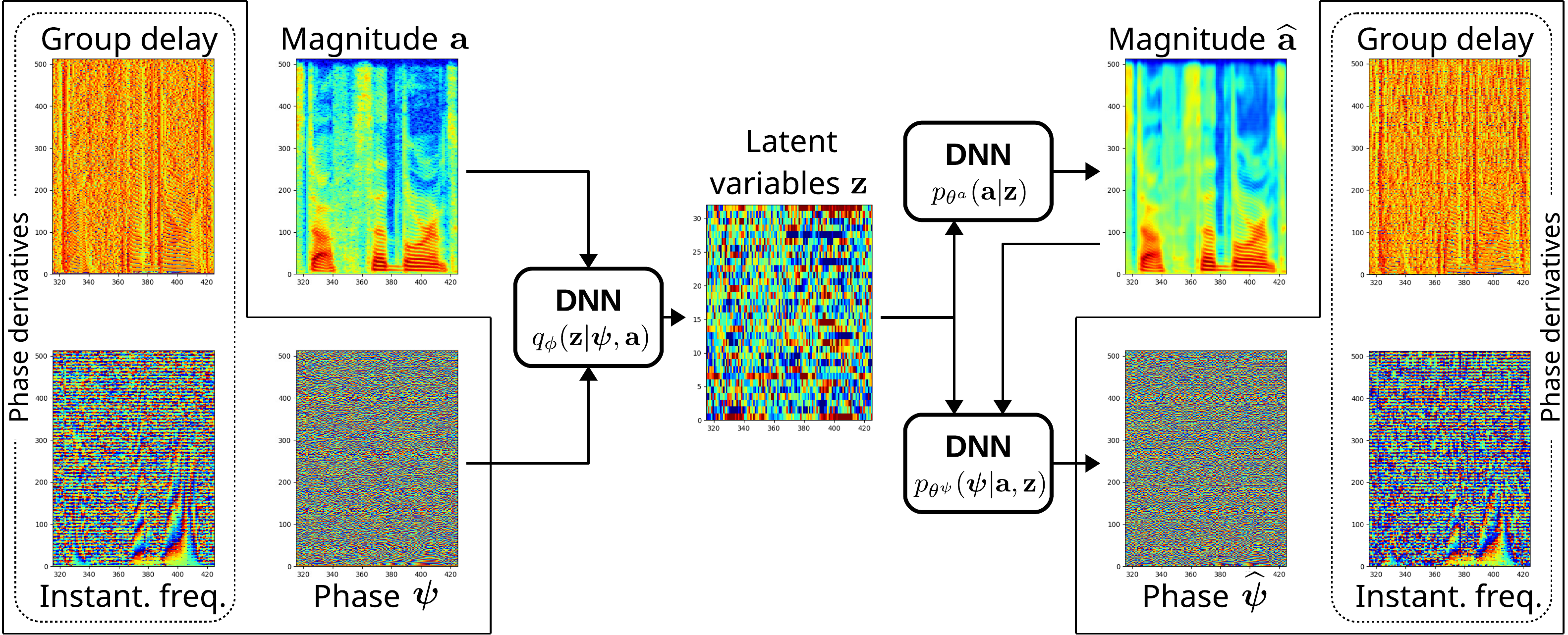}}\smallskip
	\vspace{-12pt}
	\caption{Overview of the proposed model.}
	\label{fig:overview}
\end{figure}

Motivated by the success of VAEs in modeling the power spectrogram \cite{kameoka18arxiv,seki18arxiv,bando18icassp,leglaive18mlsp,sekiguchi18apsipa} and that of DNNs in estimating the phase given the magnitude \cite{takamichi18iwaenc,takahashi18interspeech}, we aim for a joint magnitude and phase deep generative model. Following the VAE framework, we define an encoder for estimating the latent variables given the magnitude and the phase spectrograms. We then define a magnitude decoder for reconstructing the magnitude given the latent variables, and a phase decoder for reconstructing the phase given the latent variables and the reconstructed magnitude. Figure \ref{fig:overview} gives an overview of our model.
We assume a Gaussian distribution for the magnitude, and a von Mises distribution for the phase and its derivatives, i.e., the group delay (GD) and the instantaneous frequency (IF), each.
The GD is the derivative along the frequency axis, and the IF is that along the time axis. 
Thus, there is an interdependence between the phase, the GD, and the IF that has to be satisfied.
We explore different loss functions for training our models.
The experimental results show that our models could reconstruct time-domain speech.
The results also suggest that as long as the GD \textit{and} the IF estimates are good, the phase estimates are not critical for obtaining a reasonable reconstruction.


The rest of this paper is organized as follows.
Section \ref{sec:proposal} introduces the proposed approach.
Section \ref{sec:eval} presents the evaluation.
Finally, Section \ref{sec:conclusion} concludes this paper.

%

\vspace{-.75\baselineskip}
\section{Proposed method}
\label{sec:proposal}
\vspace{-.5\baselineskip}


Let $a_{f, n} \in \mathbb{R}_{\ge 0}$ and $\psi_{f, n} \in [-\pi, \pi)$ be the magnitude and the phase, respectively, of a complex-valued STFT spectrum $s_{f, n} = a_{f, n} e ^ {i \psi_{f, n}} \in \mathbb{C}$, where $f \in [1, F]$ and $n \in [1, N]$ are the frequency bin and the time frame indexes.
For a generative modeling purpose, let us now assume that for a time frame $\smash{n}$, the magnitude $\smash{\mathbf{a}_{n} = \left[ a_{1, n}, \dots, a_{F, n} \right]^\top}$ and the phase $\smash{\boldsymbol{\psi}_{n} = \left[ \psi_{1, n}, \dots, \psi_{F, n} \right]^\top}$ depend on latent variables $\smash{\mathbf{z}_{n}} \in \mathbb{R}^{D}$ with $\smash{D < F}$. Knowing that there is a generation process $\smash{p_{\theta} (\boldsymbol{\psi}_{n}, \mathbf{a}_{n} | \mathbf{z}_{n})}$ with some parameters $\theta$, we want to maximize the joint probability $\smash{p_{\theta} (\boldsymbol{\psi}_{n}, \mathbf{a}_{n})}$.

\vspace{-.75\baselineskip}
\subsection{Model formulation}
\vspace{-.25\baselineskip}

We propose to represent the joint probability between the phase $\boldsymbol{\psi}_{n}$, the magnitude $\mathbf{a}_{n}$, and the latent variables $\mathbf{z}_{n}$ as
\begin{equation}
p_{\theta} (\boldsymbol{\psi}_{n}, \mathbf{a}_{n}, \mathbf{z}_{n}) = p_{\theta^\psi} (\boldsymbol{\psi}_{n} | \mathbf{a}_{n}, \mathbf{z}_{n}) p_{\theta^a} (\mathbf{a}_{n} | \mathbf{z}_{n}) p_{\theta} (\mathbf{z}_{n}) .
\end{equation}
Note that $\boldsymbol{\psi}_{n}$ is conditioned on $\mathbf{a}_{n}$ and $\mathbf{z}_{n}$, while $\mathbf{a}_{n}$ is conditioned on $\mathbf{z}_{n}$ only.
Following the VAE framework \cite{kingma14iclr},
to approximate the posterior $\smash{p_{\theta} (\mathbf{z}_{n} | \boldsymbol{\psi}_{n}, \mathbf{a}_{n})}$,
we introduce a variational inference process $\smash{q_{\phi} (\mathbf{z}_{n} | \boldsymbol{\psi}_{n}, \mathbf{a}_{n})} \sim \smash{\mathcal{N} ( \mathbf{z}_{n} | \boldsymbol{\mu}^{q}_{n}, ( \boldsymbol{\sigma}^{q}_{n} )^2 \text{\textbf{I}} )}$ with parameters $\phi$, where $\text{\textbf{I}}$ is a $D$-dimensional identity matrix.
We also assume a simple prior $\smash{p_{\theta} ( \mathbf{z}_{n} ) \sim \mathcal{N} ( \mathbf{z}_{n} | \mathbf{0}, \text{\textbf{I}} )}$.

We then obtain a VAE with an \textit{encoder} $\smash{q_{\phi} (\mathbf{z}_{n} | \boldsymbol{\psi}_{n}, \mathbf{a}_{n})}$, and a decoder consisting of a \textit{magnitude decoder} $\smash{p_{\theta^a} (\mathbf{a}_{n} | \mathbf{z}_{n})}$ and a \textit{phase decoder} $\smash{p_{\theta^\psi} (\boldsymbol{\psi}_{n} | \mathbf{a}_{n}, \mathbf{z}_{n})}$.
Thus, there are three DNNs to be trained.
The combination of the encoder and the magnitude decoder is similar to the VAEs in \cite{bando18icassp,leglaive18mlsp}. Moreover, the phase decoder resembles the DNNs in \cite{takamichi18iwaenc,takahashi18interspeech}, that are trained in a supervised manner to estimate the phase given the magnitude spectrogram.
In our work, the above three DNNs are jointly trained in an unsupervised manner. Using the encoder, we could estimate the latent variables $\mathbf{z}_{n}$ given some observations. Most importantly, 
we could obtain the complex-valued STFTs, reconstructed using the magnitude and the phase estimated by the decoder, given $\mathbf{z}_{n}$ sampled from the simple prior \ $\smash{p_{\theta} ( \mathbf{z}_{n} )}$.

\vspace{-.75\baselineskip}
\subsection{Parameter estimation}
\label{sec:param_est}
\vspace{-.25\baselineskip}

The parameters could be jointly optimized by minimizing the negative log-likelihood (NLL) function:
\begin{align}
- \text{ln} \, p_{\theta} (\boldsymbol{\psi}_{n}, \mathbf{a}_{n}) & \!=\! - \text{ln} \int_{\mathbf{z}_{n}} p_{\theta} (\boldsymbol{\psi}_{n}, \mathbf{a}_{n}, \mathbf{z}_{n}) \, \text{d}\mathbf{z}_{n}  \nonumber \\
& \!\leq\! - \mathbb{E}_{q_{\phi} (\mathbf{z}_{n} | \boldsymbol{\psi}_{n}, \mathbf{a}_{n})} \left[ \text{ln} \,  \frac{p_{\theta} (\boldsymbol{\psi}_{n}, \mathbf{a}_{n}, \mathbf{z}_{n})}{q_{\phi} (\mathbf{z}_{n} | \boldsymbol{\psi}_{n}, \mathbf{a}_{n})} \right] \nonumber \\
& \!=\! \text{KL} [q_{\phi} (\mathbf{z}_{n} | \boldsymbol{\psi}_{n}, \mathbf{a}_{n}) || p_{\theta} (\mathbf{z}_{n}) ] \nonumber \\
& \quad \!-\! \mathbb{E}_{q_{\phi} (\mathbf{z}_{n} | \boldsymbol{\psi}_{n}, \mathbf{a}_{n})} \left[ \text{ln} \,  p_{\theta^a} (\mathbf{a}_{n} | \mathbf{z}_{n}) \right] \nonumber \\
& \quad \!-\! \mathbb{E}_{q_{\phi} (\mathbf{z}_{n} | \boldsymbol{\psi}_{n}, \mathbf{a}_{n})} \left[ \text{ln} \,  p_{\theta^\psi} (\boldsymbol{\psi}_{n} | \mathbf{a}_{n}, \mathbf{z}_{n}) \right] \! ,
\end{align}
where $\text{ln} (\cdot)$ returns the natural logarithm, $\mathbb{E} [\cdot]$ returns the expectation, and $\smash{\text{KL} [ {\triangle} || {\square} ]}$ is the Kullback-Leibler divergence from ${\square}$ to ${\triangle}$ \cite{kullback51ams}. The first term is a regularization term $\smash{\mathcal{L}^{\text{reg}}}$, the second term is a magnitude reconstruction loss $\smash{\mathcal{L}^{\text{mag}}}$, and the third term is a phase reconstruction loss $\smash{\mathcal{L}^{\text{pha}}}$.


The regularization term \cite{kingma14iclr} is expressed as
\begin{equation}
\mathcal{L}^{\text{reg}} = \frac{1}{2N} \sum_{d,n} \left( ( \mu^{q}_{d, n} )^2 + ( \sigma^{q}_{d, n} )^2 - \text{ln} ( \sigma^{q}_{d, n} )^2 - 1 \right) , \label{eq:L_reg}  \end{equation}
where $d$ is the latent variable dimension index.

The magnitude $\smash{a_{f, n}}$ is assumed to follow a Gaussian distribution with mean $\smash{\mu^{\text{mag}}_{f, n} \in \mathbb{R}_{\ge 0}}$ and variance $\smash{{( \sigma^{\text{mag}}_{f, n} )^2 \in \mathbb{R}_{\ge 0}}}$:
\begin{equation}
a_{f, n} \sim \mathcal{N} \left( a_{f, n} \middle| \mu^{\text{mag}}_{f, n}, \big( \sigma^{\text{mag}}_{f, n} \big)^2 \right) .
\end{equation}
The magnitude reconstruction loss is the NLL function:
\begin{equation}
\mathcal{L}^{\text{mag}} = \frac{1}{2N} \sum_{f, n} \left( \text{ln} \, 2 \pi \big( \widehat{\sigma}^{\text{mag}}_{f, n} \big)^2 + \frac{\big( a_{f, n} - \widehat{a}_{f, n} \big)^{2}}{\big( \widehat{\sigma}^{\text{mag}}_{f, n} \big)^2} \right) ,
\end{equation}
where the estimate $\smash{\widehat{a}_{f, n}}$ equals to the estimated mean $\smash{\widehat{\mu}^{\text{mag}}_{f, n}}$.
Additionally, we introduce a regularization term:
\begin{equation}
\mathcal{L}^{\text{var}} = \frac{1}{N} \sum_{f, n} \big( \widehat{\sigma}^{\text{mag}}_{f, n} \big)^2 ,
\end{equation}
which enforces small variances for the distribution so that obtaining small estimation errors is more emphasized. Empirically, we observed that this term is crucial when we consider more loss components, i.e., the phase-related ones.

The phase $\smash{\psi_{f, n}}$ is assumed to follow a von Mises distribution with mean $\smash{\mu^{\text{pha}}_{f, n} \in [-\pi, \pi)}$ and concentration $\smash{\kappa^{\text{pha}}_{f, n} \in \mathbb{R}_{\ge 0}}$:
\begin{equation}
\psi_{f, n} \sim \mathcal{VM} \left( \psi_{f, n} \middle| \mu^{\text{pha}}_{f, n}, \kappa^{\text{pha}}_{f, n} \right) .
\end{equation}
Several works have applied the same assumption for the phase \cite{gerkmann14tsp,takamichi18iwaenc,magron18iwaenc}.
The phase reconstruction loss is the NLL function:
\begin{equation}
\mathcal{L}^{\text{pha}} \!=\! \frac{1}{N} \sum_{f, n} \left( \text{ln} \, 2 \pi I_0 \big( \widehat{\kappa}^{\text{pha}}_{f, n} \big) - \widehat{\kappa}^{\text{pha}}_{f, n} \, \text{cos} \big( \psi_{f, n} - \widehat{\psi}_{f, n} \big) \right) , \label{eq:L_pha}
\end{equation}
where the estimate $\smash{\widehat{\psi}_{f, n}}$ is the estimated mean $\smash{\widehat{\mu}^{\text{pha}}_{f, n}}$ and $I_0 (\cdot)$ is the modified Bessel function of the first kind with order 0. 

Furthermore, we consider the phase derivatives, i.e., the group delay \cite{oppenheim09} and the instantaneous frequency \cite{boashash92ieee}. The group delay (GD) $\psi^{\text{grd}}_{f, n} \in [-\pi, \pi)$ is the phase derivative along the frequency axis:
\begin{equation}
\psi^{\text{grd}}_{f, n} = \text{wrap} ( - \psi_{f+1, n} + \psi_{f, n} ) ,
\end{equation}
and the instantaneous frequency (IF) $\psi^{\text{ifr}}_{f, n} \in [-\pi, \pi)$ is the phase derivative along the time axis:
\begin{equation}
\psi^{\text{ifr}}_{f, n} = \text{wrap} ( \psi_{f, n+1} - \psi_{f, n} ) ,
\end{equation}
where $\text{wrap} (\cdot)$ returns value in $[-\pi, \pi)$. Both derivatives capture the phase dynamics in the different axes.

We also apply the von Mises distribution on the GD $\smash{\psi^{\text{grd}}_{f, n}}$, with parameters $\smash{\mu^{{\text{grd}}}_{f, n}}$ and $\smash{\kappa^{{\text{grd}}}_{f, n}}$, and the IF $\smash{\psi^{\text{ifr}}_{f, n}}$, with parameters $\smash{\mu^{{\text{ifr}}}_{f, n}}$ and $\smash{\kappa^{{\text{ifr}}}_{f, n}}$.
We define $\mathcal{L}^{\text{grd}}$ for the GD by substituting $\smash{{\psi}_{f, n}}$, $\smash{\widehat{\psi}_{f, n}}$ and $\smash{\kappa^{\text{pha}}_{f, n}}$ in \eqref{eq:L_pha} with $\smash{{\psi}^{\text{grd}}_{f, n}}$, $\smash{\widehat{\psi}^{\text{grd}}_{f, n}}$ and $\smash{\widehat{\kappa}^{{\text{grd}}}_{f, n}}$, respectively. Similarly, we define $\mathcal{L}^{\text{ifr}}$ for the IF with $\smash{{\psi}^{\text{ifr}}_{f, n}}$, $\smash{\widehat{\psi}^{\text{ifr}}_{f, n}}$, and $\smash{\widehat{\kappa}^{{\text{ifr}}}_{f, n}}$. Note that we do not directly estimate the GD and the IF. They are derived from the estimated phase. Thus, $\mathcal{L}^{\text{grd}}$ and $\mathcal{L}^{\text{ifr}}$ can be seen as constraints, or priors, during the training.

In this paper, we do not estimate any concentration parameter and opt to set  $\smash{\widehat{\kappa}^{{\text{pha}}}_{f, n} = \widehat{\kappa}^{{\text{grd}}}_{f, n} = \widehat{\kappa}^{{\text{ifr}}}_{f, n} = \widehat{a}_{f, n} + 1}$.
This setting makes the estimation errors on $\smash{\widehat{\psi}^{\text{pha}}_{f, n}}$, $\smash{\widehat{\psi}^{\text{grd}}_{f, n}}$, and $\smash{\widehat{\psi}^{\text{ifr}}_{f, n}}$ more important when the estimated magnitude $\widehat{a}_{f, n}$ is high, and vice versa. As a comparison, another work \cite{takamichi18iwaenc} sets $\smash{\widehat{\kappa}^{{\text{pha}}}_{f, n}} = 1$.


\begin{figure}[!t]
	\hspace{-12pt}\begin{minipage}[b]{.4\linewidth}
		\centering
		\centerline{\includegraphics[height=5.9cm]{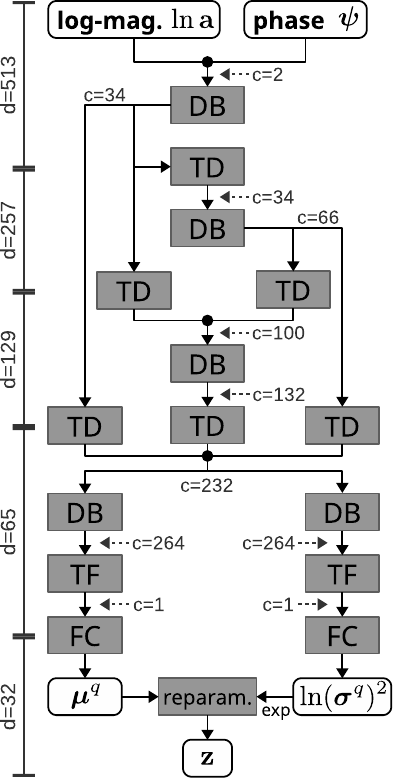}}\vspace{2pt}
		\centerline{(a) Encoder}\medskip
	\end{minipage}
	\hfill
	\begin{minipage}[b]{0.6\linewidth}
		\centering
		\centerline{\includegraphics[height=5.9cm]{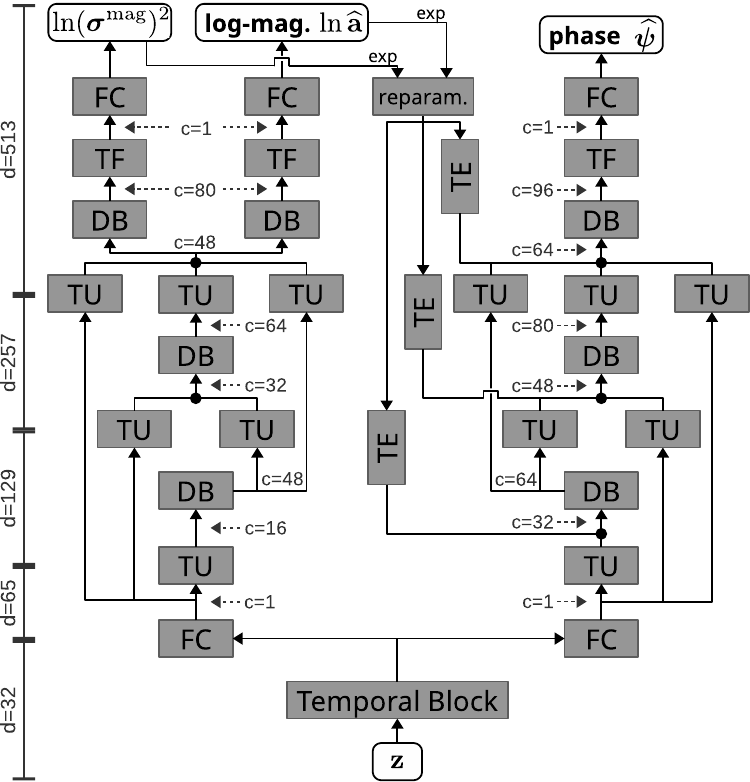}}\vspace{2pt}
		\centerline{(b) Decoder}\medskip
	\end{minipage}
	\vspace{-11pt}
	\caption{Block diagram of the model architecture. The feature map dimension is $(c, d, N)$, where $c$ is the number of channels, $d$ is the vector length for frame $n$, and $N$ is the number of frames. Black circle concatenates the input channels. The reparameterization trick \cite{kingma14iclr} is used during the training.}
	\label{fig:model}
\end{figure}

\vspace{-.75\baselineskip}
\subsection{DNN design and training}
\label{sec:model_design}
\vspace{-.25\baselineskip}


Figure \ref{fig:model} illustrates the model used in this paper. The number of frequency bins is $F \! = \! 513$ and the latent variable dimension is $D \! = \! 32$. The total number of parameters is about $1.7$ million.

Our model implementation resembles the fully convolutional DenseNets \cite{jegou17cvpr}, that combines the DenseNets \cite{huang17cvpr} and the U-Net \cite{ronneberger15miccai}. However, our model does not have skip connections between the encoder and the decoders. We employ the gated design \cite{dauphin17icml} for all convolutional layers (CLs) and the weight normalization \cite{salimans16nips}, instead of the batch normalization.

We follow the terminology in \cite{huang17cvpr, jegou17cvpr}. A Dense Block (DB) consists of $4$ two-dimensional CLs with a $3\!\times\!3$ kernel and a channel growth rate of $8$. The output channel number is the input channel number plus $4\!\times\!8$. A Transition Down (TD) consists of a $1\!\times\!1$ CL followed by $1\!\times\!1$ average pooling with an adjustable stride for reducing the vector length while keeping the channel number. A Transition Expand (TE) is similar to a TD, but it returns $16$ channels. Conversely, a Transition Up (TU) consists of a $3\!\times\!3$ transposed CL with an adjustable stride for expanding the vector length. It always returns $16$ channels. A Transition Final (TF) consists of a $1\!\times\!1$ CL to reduce the channel number to $1$. Additionally, we use a Temporal Block \cite{bai18arxiv} consisting of $4$ one-dimensional dilated CLs applied along the time frame axis. The kernel size is $3$ with dilations of $1$, $2$, $4$, and $8$ for the different layers. This block is used to capture the temporal dynamic of the latent variables. A fully-connected (FC) layer uses a leaky ReLU activation function, except when it is used as the output layer. The phase decoder outputs are in $[-\pi, \pi)$.



The model training is done in two stages. In general, the first stage aims to model the magnitude, while the second one aims to jointly model the magnitude and the phase. In the first stage, the encoder and the magnitude decoder are randomly initialized and then trained with a loss $\mathcal{L}^{\text{(M)}} = \mathcal{L}^{\text{reg}} + \mathcal{L}^{\text{mag}} + \mathcal{L}^{\text{var}}$. In the second stage, all encoder and decoders are trained with a loss $\mathcal{L}^{\text{(J)}} = \mathcal{L}^{\text{(M)}} + \mathcal{L}^{\text{(P)}}$ given the pre-trained encoder, the pre-trained magnitude decoder, and the randomly initialized phase decoder. The loss $\mathcal{L}^{\text{(P)}}$ may consist of $\mathcal{L}^{\text{pha}}$, $\mathcal{L}^{\text{grd}}$, or $\mathcal{L}^{\text{ifr}}$.
Thus, we end up with several models (see Tables \ref{tab:log_likelihood} and \ref{tab:pesq_stoi}).

\vspace{-1.75\baselineskip}
\section{Evaluation}
\label{sec:eval}
\vspace{-\baselineskip}

To evaluate the proposed approach, we consider a speech signal reconstruction task, where the latent variables are estimated given a clean utterance and then used to recreate that utterance.
The reconstructed speech quality is assessed in terms of the Mean Opinion Score (MOS) \cite{itu03pesq_moslqo}, obtained by mapping the Perceptual Evaluation of Speech Quality score \cite{itu01pesq}. The MOS ranging from 1 to 5 represents the quality ranging from bad to excellent.
Additionally, the intelligibility is assessed using the Short-Time Objective Intelligibility (STOI) score \cite{taal11taslp}.

\vspace{-1.75\baselineskip}
\subsection{Experimental settings}
\label{sec:exp_settings}
\vspace{-.5\baselineskip}

We use the speech utterances from the CHiME-4 dataset \cite{vincent16csl}, which are taken from the 5k vocabulary subset of the Wall Street Journal corpus \cite{garofalo07wsj}. All data are sampled at 16 kHz. We only consider the clean speech from the channel 5 of the simulated utterances. The training, the development, and the test sets contain 7138, 1640, and 1320 utterances, respectively. Our models are trained on the training set, validated on the development set, and evaluated on the test set.

The STFT coefficients are extracted using a Hann window with a length of 512 and a $75\%$ overlap. We then apply a 1024-point discrete Fourier transform on the windowed signals resulting in $F \! = \! 513$. The rather high zero-padding factor reveals useful features by oversampling the spectrum \cite{oppenheim09}. In our case, it exposes more evident patterns in the IF spectrogram.


The models are trained by backpropagation \cite{rumelhart86nature} with the Adam update rule whose parameters are fixed to $\alpha \! = \! 10^{-3}$, $\beta_1 \! = \! 0.9$, $\beta_2 \! = \! 0.999$, and $\epsilon \! = \!  10^{-6}$\cite{kingma15iclr}.
The update is done for every minibatch of 4096 frames, composed of 256-frame segments from 16 randomly selected utterances.
The phase of each segment is shifted with a random value sampled from $\mathcal{N} (0, 1)$ to increase its variation.
The gradient is normalized with threshold $= \! 1$ \cite{pascanu13icml}. The training is stopped after 20 consecutive epochs failed to obtain better validation error \cite{prechelt12chap}. The latest model yielding the lowest error is kept. \looseness=-1




\vspace{-.75\baselineskip}
\subsection{Experimental results}
\label{sec:exp_results}
\vspace{-.5\baselineskip}

\begin{table}[!t]
	\small
	\vspace{-\baselineskip}
	\centering
	\caption{Average log-likelihood on the test set for the different training loss functions.}%
	\label{tab:log_likelihood}
	\begin{tabularx}{\columnwidth}{c l l l l r r r r}
		\toprule
		\multicolumn{1}{c}{Model}    & \multicolumn{4}{c}{Loss function}
		& \multicolumn{1}{c}{$\smash{\widehat{\mathbf{a}}_{n}}$}
		& \multicolumn{1}{c}{$\smash{\widehat{\boldsymbol{\psi}}_{n}}$}
		& \multicolumn{1}{c}{$\smash{\widehat{\boldsymbol{\psi}}^{\text{grd}}_{n}}$}
		& \multicolumn{1}{c}{$\smash{\widehat{\boldsymbol{\psi}}^{\text{ifr}}_{n}}$} \\
		\midrule
		(M) & $\mathcal{L}^{\text{reg}}$ & \hspace{-6pt} + \hspace{1pt} $\mathcal{L}^{\text{mag}}$ & \hspace{-6pt} + \hspace{1pt} $\mathcal{L}^{\text{var}}$ & \hspace{-6pt} 
		& \textbf{1400} & -1204 & -1204 & -1204 \\
		\midrule
		(J1) & (M) & \hspace{-6pt} + \hspace{1pt} $\mathcal{L}^{\text{pha}}$ & \hspace{-6pt} & \hspace{-6pt}
		& {1366} & \textbf{-964} & -712 & -954 \\
		(J2) & (M) & \hspace{-6pt} & \hspace{-6pt} + \hspace{1pt} $\mathcal{L}^{\text{grd}}$ & \hspace{-6pt}
		& \textbf{1435} & {-1201} & \textbf{-607} & -1201 \\
		(J3) & (M) & \hspace{-6pt} & \hspace{-6pt} & \hspace{-6pt} + \hspace{1pt} $\mathcal{L}^{\text{ifr}}$ 
		& \textbf{1401} & {-1198} & -1198 & \textbf{-800} \\
		(J4) & (M) & \hspace{-6pt} + $\frac{1}{2} \mathcal{L}^{\text{pha}}$ & \hspace{-6pt} + $\frac{1}{2} \mathcal{L}^{\text{grd}}$ & \hspace{-6pt}
		& \textbf{1420} & -1053 & {-635} & -1054 \\
		(J5) & (M) & \hspace{-6pt} + $\frac{1}{2} \mathcal{L}^{\text{pha}}$ & \hspace{-6pt} &  \hspace{-6pt} + $\frac{1}{2} \mathcal{L}^{\text{ifr}}$
		& \textbf{1399} & -1191 & -1194 & {-826} \\
		(J6) & (M) & \hspace{-6pt} & \hspace{-6pt} + $\frac{1}{2} \mathcal{L}^{\text{grd}}$ & \hspace{-6pt} + $\frac{1}{2} \mathcal{L}^{\text{ifr}}$
		& \textbf{1409} & -1198 & -671 & -894 \\
		(J7) & (M) & \hspace{-6pt} + $\frac{1}{3} \mathcal{L}^{\text{pha}}$ & \hspace{-6pt} + $\frac{1}{3} \mathcal{L}^{\text{grd}}$ & \hspace{-6pt} + $\frac{1}{3} \mathcal{L}^{\text{ifr}}$
		& \textbf{1403} & -1196 & -690 & -908 \\
		\bottomrule
	\end{tabularx}
\end{table}

\begin{table}[!t]
	\small
	\vspace{-.5\baselineskip}
	\centering
	\caption{Average objective perceptual performance on the test set for the different training loss functions. The Griffin-Lim algorithm (GLA) is also considered for post-processing. }%
	\label{tab:pesq_stoi}
	\begin{tabularx}{\columnwidth}{c l l l l r r r r}
		\toprule
		\multirow{2}{*}[-1pt]{Model}    & \multicolumn{4}{c}{\multirow{2}{*}[-1pt]{Loss function}}
		& \multicolumn{2}{c}{Without GLA}   & \multicolumn{2}{c}{With GLA} \\
		\cmidrule(lr){6-7} \cmidrule(lr){8-9}
		& & & &
		& \multicolumn{1}{c}{MOS}   	& \multicolumn{1}{c}{STOI}
		& \multicolumn{1}{c}{MOS}   	& \multicolumn{1}{c}{STOI} \\
		\midrule
		(M) & $\mathcal{L}^{\text{reg}}$ & \hspace{-6pt} + \hspace{1pt} $\mathcal{L}^{\text{mag}}$ & \hspace{-6pt} + \hspace{1pt} $\mathcal{L}^{\text{var}}$ & \hspace{-6pt} 
		& 1.96 & 0.690 & 3.97 & \textbf{0.792} \\
		\midrule
		(J1) & (M) & \hspace{-6pt} + \hspace{1pt} $\mathcal{L}^{\text{pha}}$ & \hspace{-6pt} & \hspace{-6pt}
		& {3.34} & {0.770} & 3.83 & \textbf{0.787} \\
		(J2) & (M) & \hspace{-6pt} & \hspace{-6pt} + \hspace{1pt} $\mathcal{L}^{\text{grd}}$ & \hspace{-6pt}
		& 2.18 & {0.734} & {4.00} & \textbf{0.795} \\
		(J3) & (M) & \hspace{-6pt} & \hspace{-6pt} & \hspace{-6pt} + \hspace{1pt} $\mathcal{L}^{\text{ifr}}$ 
		& 2.51 & {0.702} & 3.86 & \textbf{0.789} \\
		(J4) & (M) & \hspace{-6pt} + $\frac{1}{2} \mathcal{L}^{\text{pha}}$ & \hspace{-6pt} + $\frac{1}{2} \mathcal{L}^{\text{grd}}$ & \hspace{-6pt}
		& \textbf{3.71} & \textbf{0.786} & \textbf{4.04} & \textbf{0.792} \\
		(J5) & (M) & \hspace{-6pt} + $\frac{1}{2} \mathcal{L}^{\text{pha}}$ & \hspace{-6pt} &  \hspace{-6pt} + $\frac{1}{2} \mathcal{L}^{\text{ifr}}$
		& 2.39 & 0.690 & 3.89 & \textbf{0.790} \\
		(J6) & (M) & \hspace{-6pt} & \hspace{-6pt} + $\frac{1}{2} \mathcal{L}^{\text{grd}}$ & \hspace{-6pt} + $\frac{1}{2} \mathcal{L}^{\text{ifr}}$
		& 3.54 & {0.777} & 3.90 & \textbf{0.789} \\
		(J7) & (M) & \hspace{-6pt} + $\frac{1}{3} \mathcal{L}^{\text{pha}}$ & \hspace{-6pt} + $\frac{1}{3} \mathcal{L}^{\text{grd}}$ & \hspace{-6pt} + $\frac{1}{3} \mathcal{L}^{\text{ifr}}$
		& 3.13 & {0.766} & 3.86 & \textbf{0.789} \\
		\bottomrule
	\end{tabularx}
\end{table}

Tables \ref{tab:log_likelihood} and \ref{tab:pesq_stoi} show the experimental results for the different loss functions on the test set. Table \ref{tab:log_likelihood} shows the average log-likelihood (LL). It is obtained by computing $-\mathcal{L}^{\text{mag}}$, $-\mathcal{L}^{\text{pha}}$, $-\mathcal{L}^{\text{grd}}$, and $-\mathcal{L}^{\text{ifr}}$ for each utterance and then averaging the results. These LL values reflect the estimation accuracy. Table \ref{tab:pesq_stoi} shows the objective perceptual performance.
The magnitude and the phase are always estimated using our models, except for the model (M) where the phase is sampled randomly from a uniform distribution. This model (M) is obtained from the first training stage only and acts as the baseline. In addition, we consider the Griffin-Lim algorithm (GLA) \cite{griffin84tassp} as post-processing. It is done for 100 iterations. Boldface numbers show the best performance for each column, taking into account the 95\% confidence interval. A higher value is better for all metrics.


\vspace{-\baselineskip}
\subsection{Discussion}
\label{sec:exp_discussion}
\vspace{-.5\baselineskip}

Table \ref{tab:log_likelihood} shows that the good magnitude reconstruction achieved by the model (M) could be preserved by the other models in most cases.
Thus, we could focus on observing the estimation of the phase and its derivatives.
The model (J1) shows that a good phase estimation naturally provides fair estimates of the derivatives.
Conversely, the models (J4), (J5), and (J7) have better estimates of either or both of GD and IF, but worse phase estimate, than the model (J1).
It suggests that the optimization of the phase derivatives strongly drives the overall optimization. Thus, a more elaborate weighting scheme might be useful.

Let us now observe Tables \ref{tab:log_likelihood} and \ref{tab:pesq_stoi} together.
The model (J4) provides the best performance.
It suggests that minimizing $\mathcal{L}^{\text{grd}}$ is useful.
The models (J1), (J4), (J6), and (J7) provide fair performance and all of them have a good GD estimation. However, the model (J2) shows that a good GD alone is not enough.
Interestingly, the models (J6) and (J7) provide reasonable performance although the phase estimation is poor. It might suggest that estimating the absolute phase value is not critical, and capturing the phase interdependence on the frequency \textit{and} the time axes is sufficient. Additionally, the GLA iterations effectively improve both the quality and the intelligibility. The performance of our models without the GLA is still below that of the GLA with random initial phase. However, our method does not need any iteration.

\begin{figure}[!t]
	\vspace{-4pt}
	\small
	\centering
	\begin{minipage}[b]{.425\linewidth}
		\centering
		\centerline{{\includegraphics[height=6.2cm,trim={0 0 110pt 0},clip]{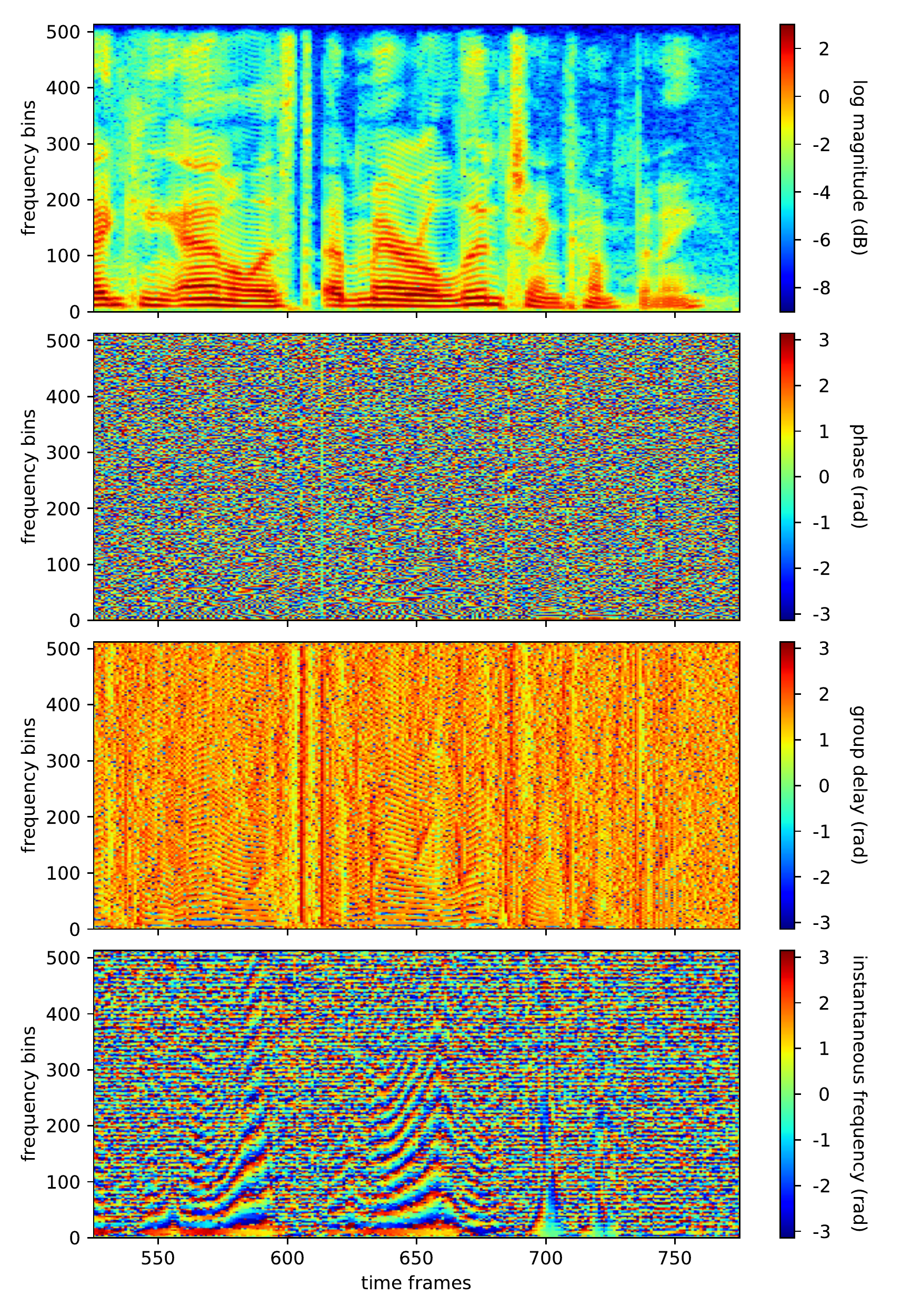}}}
	\end{minipage}
	\begin{minipage}[b]{0.425\linewidth}
		\centering
		\centerline{{\includegraphics[height=6.2cm,trim={48pt 0 0 0},clip]{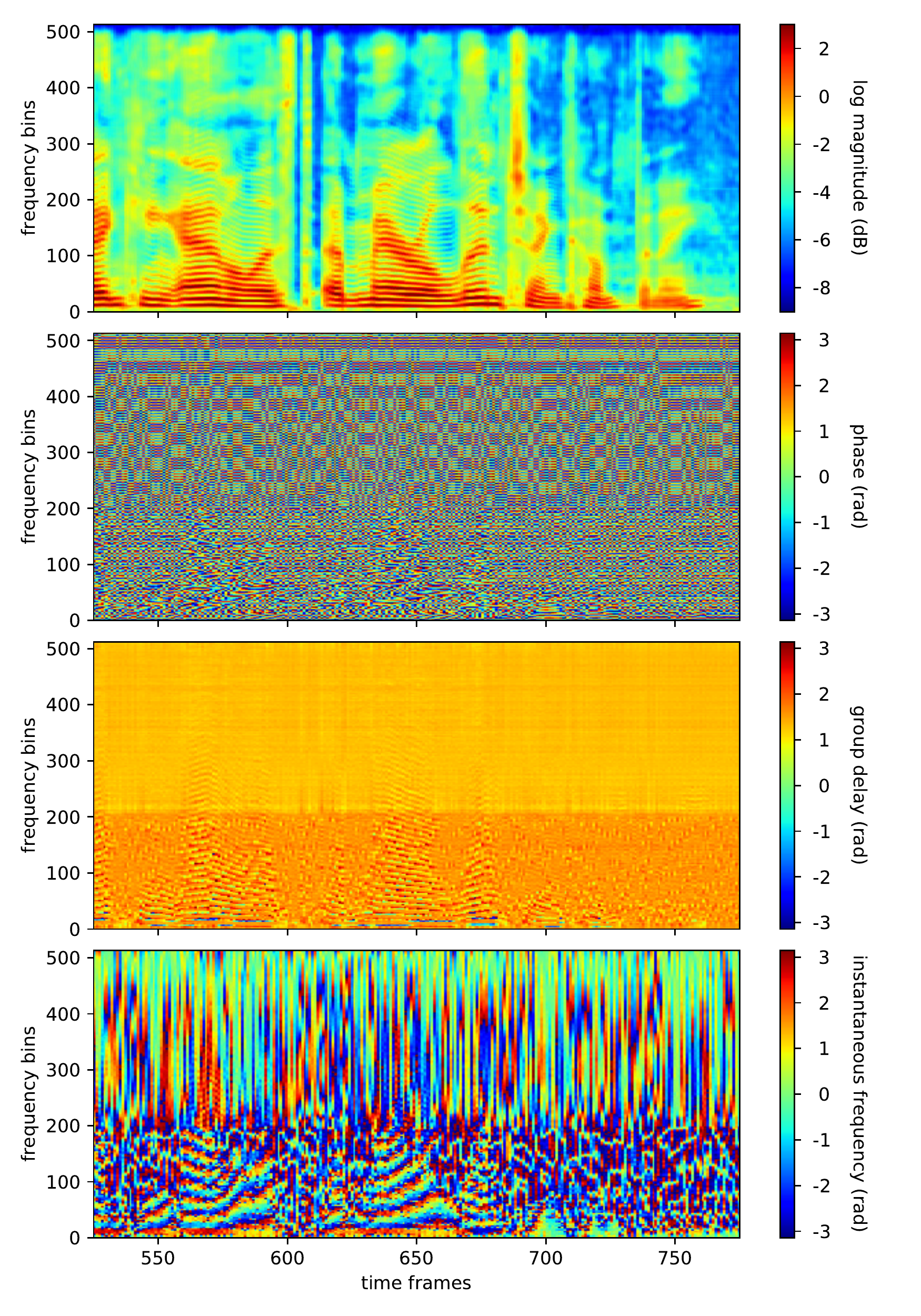}}}
	\end{minipage}
	\vspace{-12pt}
	\caption{The left and the right columns show spectrogram examples of a true speech and its reconstruction using the model (J4), respectively. From top to bottom, we show the log-magnitude, the phase, the group delay, and the instantaneous frequency spectrograms. The utterance is \small{\texttt{F05\_440C020I\_PED}} from the  set \small{\texttt{et05\_ped\_simu}}.}
	\label{fig:spectrograms}
\end{figure}

Figure \ref{fig:spectrograms} shows spectrogram examples of a speech segment and its reconstruction. The log-magnitude spectrograms show that the model reconstructs the harmonic structures well, although those for above the frequency bin 200 tend to be unclear. The phase and the phase derivative spectrograms clearly show that there are still opportunities for further improvement. The estimated spectrograms resemble the true ones only for the lower frequency bands. This might be an impact of associating the von Mises concentration parameters to the estimated magnitude. Therefore, the estimation of those parameters should be explored further.
Audio samples are available online\footnote{{Demo webpage: \scriptsize{\url{https://aanugraha.gitlab.io/demo/icassp19}}}}.

\vspace{-.75\baselineskip}
\section{Conclusion}
\label{sec:conclusion}
\vspace{-.75\baselineskip}


We proposed a deep generative model for jointly modeling the magnitude and the phase of STFT. We took into account the phase derivatives, i.e., the group delay and the instantaneous frequency.
We found that good phase derivative estimates are sufficient to provide a fair speech quality.
However, we also found that the phase derivative optimization strongly drives the overall optimization and thus, a more elaborate weighting scheme might be required. 
Additionally, future work includes incorporating the estimation of the von Mises concentration parameters and utilizing the proposed models for downstream tasks, e.g., speech enhancement and audio source separation.



\cleardoublepage

\begingroup
\def\baselinestretch{.85}\let\normalsize\small\normalsize
\bibliographystyle{IEEEbib}
\bibliography{refs}
\endgroup

\end{document}